\documentstyle[sprocl]{article}

\input epsf

\def\be{\begin{equation}}
\def\ee{\end{equation}}
\def\bea{\begin{eqnarray}}
\def\eea{\end{eqnarray}}

\begin{document}

\title{Precision Electroweak Measurements at the SLC :
Overview and Perspective}

\author{P.C. Rowson}

\address{Stanford Linear Accelerator Center, mail stop 96, 
Stanford,\\CA 94309, USA\\E-mail: rowson@slac.stanford.edu}


\maketitle\abstracts{ Preliminary SLD electroweak results 
based on essentially the complete 550K Z dataset are presented and interpreted,
and some historical background is provided.}

\section{Introduction}

The recent run (1997-98) of the Stanford Linear Collider (SLC) 
was the most productive to date :
approximately 350K Z bosons were detected, compared to 210K
for the entire program from 1992-96, at peak luminosities
of $3 \times 10^{30} {\rm cm^{-2} s^{-1}}$, nearly a
factor of three improvement compared to the
best previous results.  Nevertheless, LEP enjoys a
28 : 1 advantage in statistics, and it is only due to the
unique features of SLC operation that the SLD experiment is able
to contribute several state-of-the-art electroweak and b-physics
measurements.   These well-known features are :
\begin{itemize}
\item High (75\%), precisely measured (${{\delta \cal P} \over{P}} \sim 0.5\%$) 
longitudinal e- polarization.
\item A small and stable $e^+e^-$ luminous region (1.5 by 0.8 by 700 $\mu m$)
and a uniquely precise CCD-based vertex detector 
(I.P. determined to 4 by 4 by 30 $\mu m$).
\end{itemize}
In what follows, recent electroweak results will be summarized, some historical
background provided, and implications of the data will be discussed. 

\section{The Electroweak Observables}

The polarized differential cross section at the Z pole is given by :

$$ {d\sigma \over {d cos\theta}} \sim (1 - {\cal P}_e A_e)(1 + cos^2\theta) 
   + 2A_f(A_e - {\cal P}_e)cos\theta, $$
where the parity violating asymmetries in terms of the
vector and axial vector NC couplings for fermion flavor $f$ 
are $A_f = {2v_f a_f \over {v_f^2 + a_f^2}}$.
The polarized $e^-$ beam at the SLC allows for the isolation of the
initial state ($A_e$) and final state ($A_f$) asymmetries.  The initial state
couplings are determined most precisely via the left-right Z production asymmetry

$$ A_{LR}^0 = {1 \over {\cal P}_e} 
   {\sigma_L - \sigma_R \over {\sigma_L + \sigma_R}} = A_e , $$
while the left-right-forward-backward asymmetry for the 
final state flavor $f = b,c,s,e,\mu ,\tau$  

$$ A_{LRFB} =  
        {(\sigma_{LF} - \sigma_{LB}) - (\sigma_{RF} - \sigma_{RB}) 
  \over {(\sigma_{LF} + \sigma_{LB}) + (\sigma_{RF} + \sigma_{RB})}} 
  = {3 \over{4}} {\cal P}_e A_f , $$
determines the final state couplings.  

The $A_{LR}$ measurement is unique among all electroweak precision
measurements in that no efficiency or acceptance effects enter, and no
significant final state identification is required.
Due to the
extensively crosschecked high precision polarimetry, 
the total systematic error ($\sim 0.75\%$)
ensures that the result is statistically dominated (stat. error $\sim 1.3\%$).
The quantity $A_{LR}^0$ provides by far the 
most precise determination of
$sin^2\theta_W^{eff}$ presently available, and rivals
the 4 experiment CERN
average, without recourse to the assumption
of lepton/hadron universality inherent in the most precise
technique from LEP ($A_{FB}(b)$).

The significance of the $A_{LRFB}$ measurement, while not as precise as
$A_{LR}$, is that it provides the only direct measurement of the important
parameter $A_b$ (and the charm, strange and muon analogs as well), which 
can only be obtained indirectly from unpolarized asymmetries. 
While the weak mixing angle measurements are particularly sensitive to
vacuum polarization loop effects (and hence to the Higgs mass),
$A_b$ is instead 
affected by corrections at the Zb${\rm \overline{b}}$ vertex.
In the context of the Minimal Standard Model (MSM), these
vertex corrections are insensitive to $M_{Higgs}$ and hence $A_b$
has an unambiguous predicted value (compared to experimental precision).
The combination of independent measurements of $A_e$ and $A_b$
is therefore a powerful test of the MSM.

In addition, measurements of the hadronic partial width ratios
$R_b$ and $R_c$, which are best measured at LEP and SLD, respectively,
have become precisely known.  In particular, $R_b$ is interesting 
due to high precision (0.4\% in the world average), and the fact that 
it provides a nicely complimentary measurement to $A_b$ : $A_b$ is primarily
sensitive to {\em right}-handed NC b couplings, while $R_b$ is most sensitive
to the {\em left}-handed sector.

\section{Remarks on High Precision }

The unique precision of $A_{LR}$ is a centerpiece of the SLD
program, but the extensive
crosschecks which have bolstered our confidence in this
measurement are not so well known and are briefly reviewed here.

In the early years (1992-95), a number of dedicated accelerator
experiments were performed to establish the integrity of the 
polarimetry, in particular 1) the $e^-$ bunch helicity transmission was verified
by setting up a current/helicity correlation in the SLC,
2) medium precision M\o ller and Mott polarimeters confirmed
the high precision Compton polarimeter result to $\sim 3\%$.  In addition,
the advent of spin manipulation via ``spin bumps" in the SLC arcs
allowed us to minimize the spin chromaticity ($d{\cal P}/dE$)
which helped reduce a resulting polarization correction from $>1\%$ in 1993 to $<0.2\%$ by 1995.

Since 1997, two additional detectors of the Compton scattered photons
(the Compton $e^-$ are seen in the primary device), with rather different
systematics, presently confirm our overall polarization scale to within $0.5\%$.
Most recently, two longstanding questions were answered : 1) A dedicated experiment using
the End Station A fixed target polarimeter confirmed that accidental $e^+$ polarization
is consistent with zero ($-0.02\pm0.07\%$), 2) A short resonance scan was used to calibrate the 
SLC energy spectrometers against $M_Z$, verifying their accuracy on $E_{cm}$
to about 40 MeV and leading to an estimate of induced systematic
error of $\sim 0.5\%$. 
\footnote{This result was somewhat inflated 
by instrumental problems during the scan, 
compared to our prior estimate of $\sim 0.4\%$, but
remains at or below the polarimeter uncertainty.}

In summary, several years of instrumental work and crosschecks, supplemented by
extensive accelerator based tests, have answered
a large number of detailed questions, from the most fundamental to the
fairly obscure.  The high precision of $A_{LR}$ is now very well established. 

\section{Results and Interpretation}

The preliminary results are given below
(with the exception of kaon tagging for $A_b$,
and the latest $\sim 100K$ events for $R_b$,
these results are based on the entire
1992-1998 SLD data set). \cite{mor99}

\begin{table}[ht]
\caption{SLD electroweak results.\label{tab:exp}}
\vspace{0.2cm}
\begin{center}
\footnotesize
\begin{tabular}{|c|c|c|l|}
\hline
{ Observable } & {Prelim. Result} &
{$sin^2\theta_W$} & {comments}\\
\hline
\hline
{ $A_{LR}$ } & {$0.1504\pm0.0023$} &
{$0.23109\pm0.00029$} & {Incl. SLD leptonic result}\\
\hline
\hline
{ $A_{e}$ } & {$0.1504\pm0.0072$} &
{} & {(The LEP leptons only}\\
{ $A_{\mu}$ } & {$0.120\pm0.019$} &
{($0.2317\pm0.0008$)} & {result for $sin^2\theta_W$ :}\\
{ $A_{\tau}$ } & {$0.142\pm0.019$} &
{} & {$0.23153\pm0.00034$)}\\
\hline
{ $A_c$ } & {$0.634\pm0.027$} &
{} & {(LEP : $0.634\pm0.040$)}\\
{ $A_b$ } & {$0.898\pm0.029$} &
{} & {(LEP : $0.887\pm0.021$)}\\
\hline
{ $R_c$ } & {$0.169\pm0.006$} &
{} & {These observables are}\\
{ $R_b$ } & {$0.2159\pm0.0020$} &
{} & {consistent with the MSM.}\\
\hline
\end{tabular}
\end{center}
\end{table}

A few comments are in order : 
\begin{itemize}
\item Final errors will be $\sim \pm 0.00025$ for $sin^2\theta_W^{eff}$,
and $\sim \pm 0.022$ for $A_b$, mainly due to improved systematics. 
\item Additional $sin^2\theta_W^{eff}$ information
derives from the left-right asymmetry in the lepton sample
(the dominant $A_{LR}$ result is from a hadronic sample).  
\item The $A_b$ result is just over one sigma away from the SM prediction
(0.935), but if $A_b$ is deduced from the LEP $A_{FB}(b)$ measurement, the
SLD/LEP combined result is 2.6 sigma low.
\item The SLD $sin^2\theta_W^{eff}$ result is nicely 
consistent with lepton-based 
results from LEP (0.8 sigma), a situation that has
held stably since 1995, while the $A_{FB}(b)$ dominated LEP hadronic
average differs from the lepton based result by 2.2 sigma.
\end{itemize}
The difficulty seen with the b-flavor results may be a
statistical fluctuation or due to analysis bias, or may point to the
intruiging possibility of an anamoly in b-NC couplings, in particular,
the right-handed coupling, a situation
which is difficult to motivate theoretically 
(the $R_b$ world average is consistent with the MSM).
In either of the later two cases, the b-hadron based $sin^2\theta_W^{eff}$ 
result is called into question - for now we feel it is reasonable to
perform our electroweak fits using the lepton-based weak mixing angle
average.

We first work within the framework of the MSM - Figure 1 shows the
result of separate fits to the Higgs mass using the lepton-based 
$sin^2\theta_W^{eff}$ results from SLD and LEP, the $M_W$ results from
LEP II and the Tevatron, and for comparison the result using the 
$A_{FB}(b)$ based weak mixing angle measurement from LEP.  
\footnote{The $\alpha(M_Z^2)$ of Kuhn etal. is used for the 
fits discussed here - the Jegerlehner etal. value used by the LEP EWWG
yields a $\chi^2$ minimum about 30 GeV lower, but due to larger
errors, 
provides about the same 95\% confidence upper bound.\cite{alpha}}
The $M_W$ measurements seem to be confirming the very low Higgs
mass favored by $A_{LR}$.  It is also noteworthy that even when $M_W$
precision reaches $\pm 30$ MeV (presently
$\pm 44$ MeV), the strongest contraints will still be
coming from $sin^2\theta_W^{eff}$.  The key to this enterprise
is that improved $\alpha(M_Z^2)$ determinations are becoming available \cite{alpha},
with further improvements expected from new low energy R data 
\footnote{Improved data for the critical 2-5 GeV region is already available
from BES, with an eventual factor of two improvement in precision 
expected in this region. \cite{DPF} }.  To fully exploit higher precision 
in $\alpha(M_Z^2)$ will
also require the expected FNAL Run II
improvements in $\delta m_{top}$, from the present 5 GeV to below 3 GeV.

\begin{figure} [h]
\begin{minipage}{5.8cm}
\leavevmode\centering
\epsfxsize=5.8cm
\epsfbox{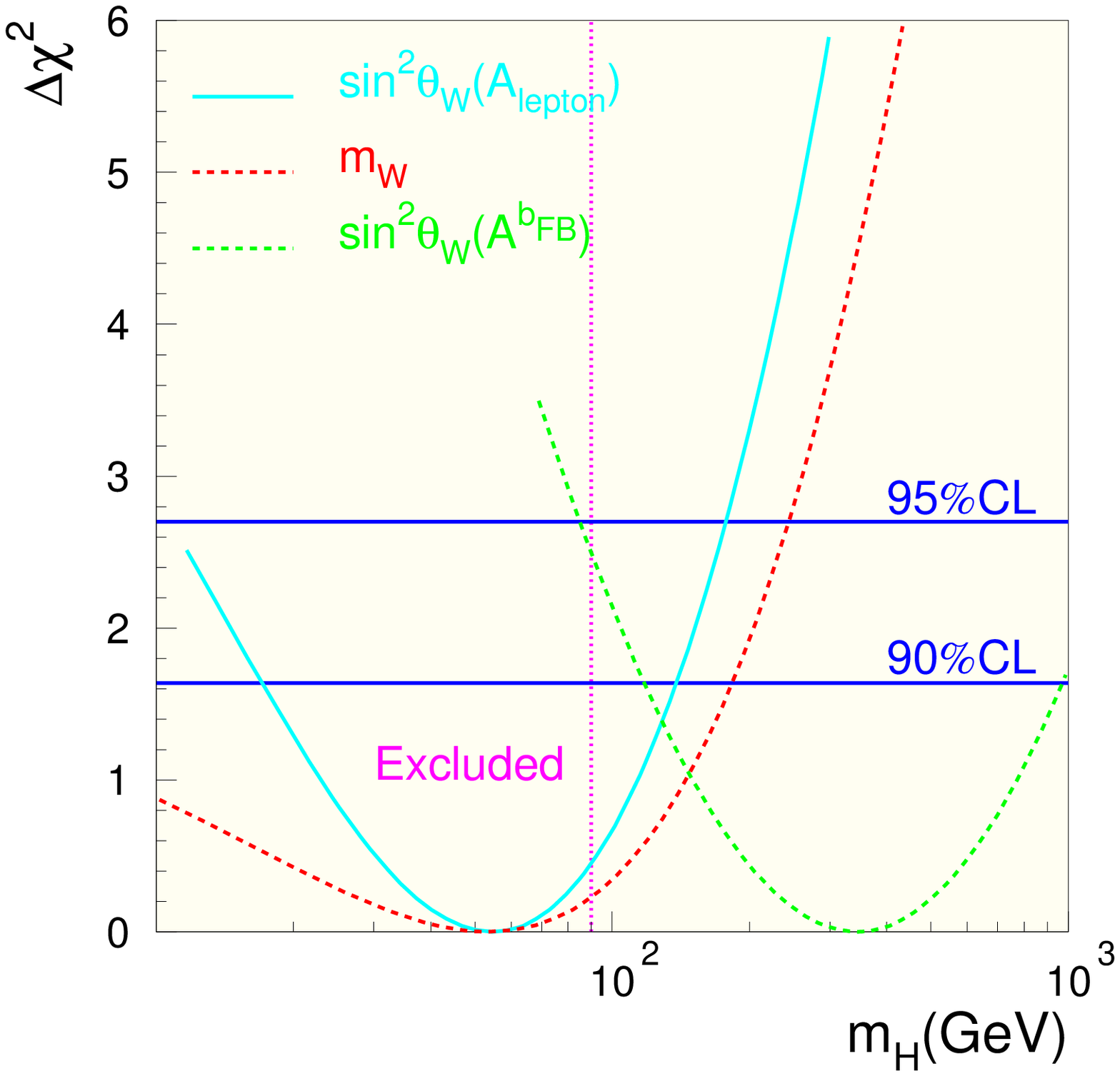}
\caption{Selected MSM fits.}
\end{minipage}
\hfill
\begin{minipage}{5.8cm}
\leavevmode\centering
\epsfxsize=5.8cm
\epsfbox{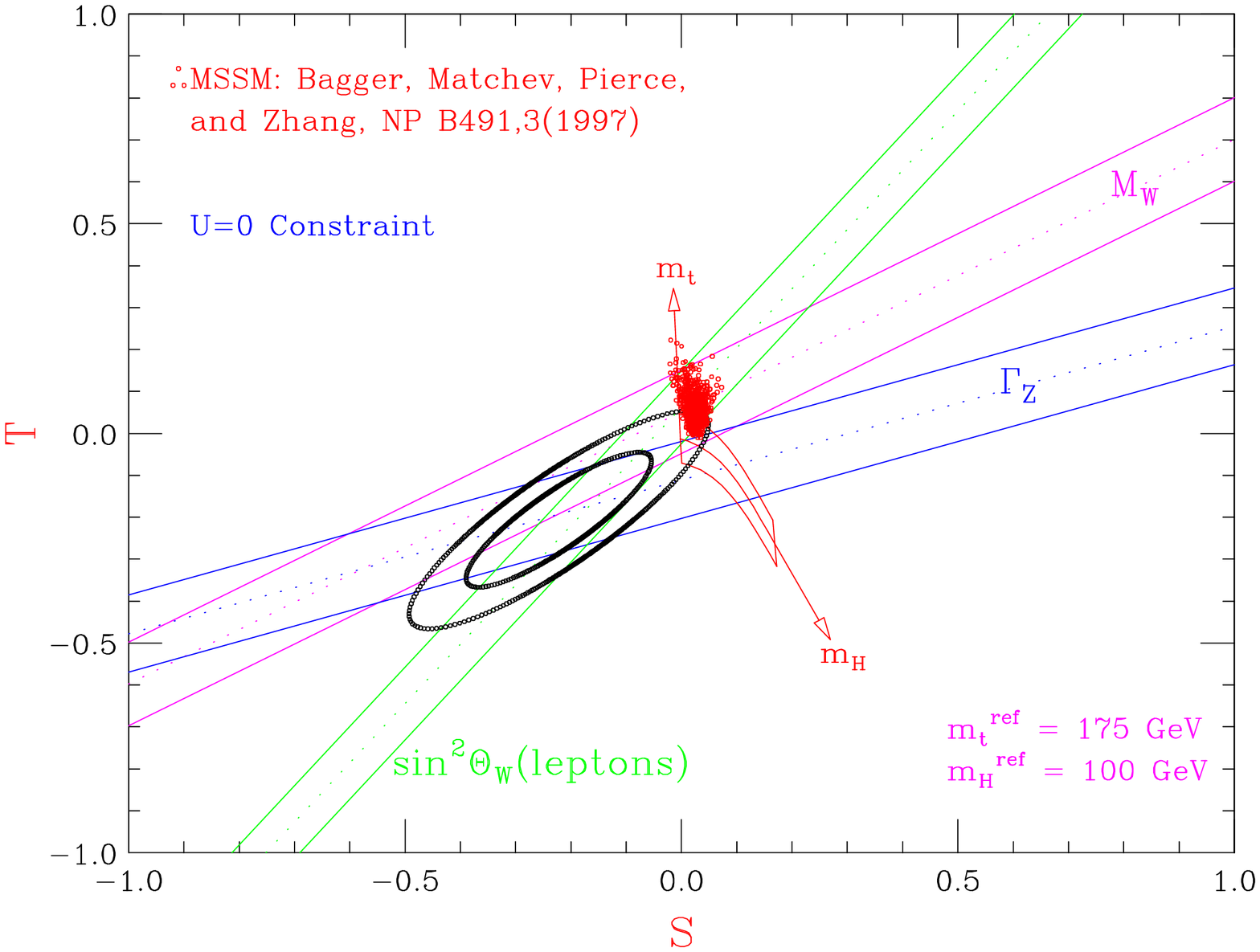}
\caption{Global S,T fit.}
\end{minipage}
\end{figure}
A more general approach employs a fit to the S,T,U parameters \cite{ST},
which encompass a broad class of models dominated by oblique radiative
effects, including supersymmetric models.  We perform a global fit to all
the world's electroweak data, including $sin^2\theta_W^{eff}$, $M_W$,
Z-width and leptonic BRs, DIS-$\nu$ scattering, and atomic parity violation,
but excluding the heavy quark results from LEP and SLD (as these may
be showing significant vertex corrections).  Figure 2 shows
the 68\% and 90\% fit ellipses and the contributions
from the three most precise inputs to the fit.
The MSM allows 
the banana-shaped shaped region, whose size is limited by the present
FNAL top mass errors, and the LEP II direct Higgs search bounds (a
value of 98 GeV for the combined result is used here).     
Also shown are a collection of points sampled from the 5-parameter
space of the Minimal Supersymetric Model (MSSM).  It is evident
how light Higgs masses, and hence the MSSM, are presently
favored (in particular by $sin^2\theta_W^{eff}$ and $M_W$).  It is also
intriguiging how $sin^2\theta_W^{eff}$ has begun to place limits
on MSSM parameters, and with improved precision could play a role in untangling
ambiguous Higgs observations at the LHC.


\section*{References}
\begin{thebibliography}{99}
\bibitem{mor99}N. DeGroot, {\em Recontres de Moriond}, March 13-20, 1999.

\bibitem{alpha}For a review of $\alpha(M_Z^2)$ results see F. Jegerlehner, 
hep-ph/9901386.

\bibitem{DPF}D. Kong, DFP'99, January 5-9, 1999.

\bibitem{ST}M. Peskin and T. Takeuchi, {\em Phys.Rev.}D46:381-409,1992.

\end{thebibliography}
\end{document}